\documentclass[amsmath,twocolumn,superscriptaddress,prx,aps]{revtex4-1} %nofootinbib
\usepackage{amsthm,amsfonts,graphicx,verbatim, color}
\usepackage[singlelinecheck=false]{caption}
\usepackage{bm}% bold math
\usepackage[utf8]{inputenc}
\usepackage{ulem}
\let\oldmarginpar\marginpar
\renewcommand\marginpar[1]{\-\oldmarginpar[\raggedleft\footnotesize #1]%
{\raggedright\footnotesize #1}}

\newcommand{\be}{\begin{equation}}
\newcommand{\ee}{\end{equation}}
\newcommand{\bea}{\begin{eqnarray}}
\newcommand{\eea}{\end{eqnarray}}

\renewcommand{\epsilon}{\varepsilon}
\renewcommand{\vec}[1]{{\bf #1}}

\def\beq{\begin{equation}}
\def\eeq{\end{equation}}
\def\bea{\begin{eqnarray}}
\def\eea{\end{eqnarray}}

\begin{document}

\title{Critical localization with Van der Waals interactions}
\author{Rahul Nandkishore}
\affiliation{Department of Physics and Center for Theory of Quantum Matter, University of Colorado at Boulder, Boulder CO 80309, USA}
\affiliation{Department of Physics, Stanford University, Stanford, CA 94305, USA}

\begin{abstract}
    I discuss the quantum dynamics of strongly disordered quantum systems with critically long range interactions, decaying as $1/r^{2d}$ in $d$ spatial dimensions. I argue that, contrary to expectations, localization in such systems is {\it stable} at low orders in perturbation theory, giving rise to an unusual `critically many body localized regime.' I discuss the phenomenology of this critical MBL regime, which includes distinctive signatures in entanglement, charge statistics, noise, and transport. Experimentally, such a critically localized regime can be realized in three dimensional systems with Van der Waals interactions, such as Rydberg atoms, and in one dimensional systems with $1/r^2$ interactions, such as trapped ions. I estimate timescales on which high order perturbative and non-perturbative (avalanche) phenomena may destabilize this critically MBL regime, and conclude that the avalanche sets the limiting timescale, in the limit of strong disorder / weak interactions.
\end{abstract}
\maketitle

Non-equilibrium many body quantum dynamics has aroused intense interest over the past decade. A cornerstone of our understanding is the phenomenon of {\it many body localization} (MBL) (see \cite{MBLARCMP, mblrmp} for reviews), by which strongly disordered quantum systems can fail to equilibrate, and can realize qualitatively new kinds of quantum phases of matter. Most work on MBL has focused on systems with purely {\it short range} interactions. However, long range interactions, which decay as a power law of distance, are ubiquitous in nature, ranging from Coulomb interactions between charges, to dipolar interactions between spins, to Van der Waals interactions between molecules and Rydberg atoms. What happens to MBL in the presence of long range interactions? 

The lore on MBL in the presence of long range interactions is built on three results. Firstly, for {\it non-interacting} problems with long range hopping, classic results \cite{pwa, LevitovDipoles} establish that if the hopping matrix element decays at long distance as $1/r^{\alpha}$, then systems can be localized as long as $\alpha > d$, where $d$ is the spatial dimension. For $\alpha < d$ the system is thermal, and the critical case is thermal but not diffusive, supporting instead subdiffusive transport \cite{LevitovDipoles}. Secondly, a generalization of this argument to {\it interacting} systems \cite{Burin, YaoDipoles} suggests that for systems with long range two-body interactions decaying as $1/r^{\beta}$, localization is perturbatively stable as long as $\beta > 2d$, but is perturbatively unstable if $\beta < 2d$ (although specific counterexamples are known \cite{LRMBL, Akhtar}), with explicit relaxation rates having been computed in \cite{NGdipoles}. However, what generically happens in the critical case $\beta = 2d$ has never been resolved, and this case is experimentally relevant to both Van der Waals interactions in three spatial dimensions, and to trapped ions in one spatial dimension, as well as being an important theoretical point of principle. Intuition from the case of long range hopping, and also from studies of (de)localization at critical points \cite{CPdelocalization} would suggest that the critical case should be delocalized, but might possibly have some unusual features. Thirdly, non-perturbative arguments \cite{avalanche} suggest that {\it any} power law interaction should produce an `avalanche instability.' I will make the conservative (and increasingly standard) assumption that this `avalanche instability' destroys the MBL {\it phase}. Nevertheless, the timescale associated to the avalanche instability is superpolynomially long in disorder strength \cite{ehudavalanche, gopalakrishnanhuse}, and upto this long timescale (which could be longer than experimental timescales), the quantum dynamics exhibits an MBL {\it regime} which will be my focus herein. 

In this work I show that, contrary to expectations, the case of `critically' long ranged interactions ($\beta = 2d$) admits of a perturbatively stable MBL regime. However the localization is of an unusual `critical' kind, with sharp few body resonances uniformly distributed in logarithmic lengthscale. I discuss the distinctive phenomenological signatures of this `critically MBL' regime. I also discuss the timescale upto which it is expected to be stable, which (I argue) is set by the avalanche instability. 

This article is structured as follows. I begin by reviewing the basic arguments for critically-long-range hopping problems \cite{LevitovDipoles}. I then generalize this approach to critically-long-range interactions, and demonstrate that the localized phase is (critically) stable at low orders in perturbation theory. I discuss the phenomenology of the resulting critically localized regime, before concluding with a discussion of timescales upto which the regime may be expected to survive, which I argue are set by the avalanche.  I work throughout on the lattice, avoiding the complications inherent with analyses in the continuum \cite{mblcontinuum1, mblcontinuum2, mblcontinuum3}. 

I begin by reviewing at a cartoon level the behavior of non-interacting systems with long range hopping, since this introduces the basic approach I will employ. I start by switching off the long range component of the hopping, and assume that in this limit the system consists of a set of Anderson localized wavefunctions $|\alpha\rangle$, with eigenenergies $\epsilon_{\alpha}$ drawn from a distribution of width $W$. Now reintroduce the long range hopping, so that the Hamiltonian becomes $H = \sum_{\alpha} \epsilon_{\alpha} c^{\dag}_{\alpha} c_{\alpha} + \frac{t}{|r_{\alpha}-r_{\beta}|^d} c^{\dag}_{\alpha} c_{\beta}$. Since we are concerned primarily with the long range tail of the hopping, the finite size of the wavefunctions $\alpha$ is not important. I treat the long range hopping perturbatively, using the method of logarithmic shells employed in \cite{Levitov89}. For a given excitation, and given a lengthscale $R$ and an integer $k$, in the logarithmic shell at distance $r$ satisfying $R 2^k < r < R 2^{k+1}$ there are $\sim R^d$ states to which the excitation can hop, with a minimum level spacing of $\sim W/R^d$. Meanwhile, the matrix element for the hopping is $t/R^d$. Hopping can be resonant if and only if the matrix element exceeds the level spacing. This ratio is $\sim t/W = \lambda$ which is crucially independent of both $R$ and $k$. If we assume $\lambda \ll 1$ (strong disorder), it then follows that for the $k^{th}$ logarithmic shell, the probability of a resonant hop is $\sim \lambda \ll 1$. This has two consequences. Firstly, it is very unlikely that one finds a resonance at the same energy scale between $n$ sites with $n>2$ - the most common resonances are between {\it pairs}, and these resonances are sparse in `logarithmic' space. Secondly, one almost surely finds a resonance at {\it some} lengthscale, since the probability of not finding {\it any} resonance in the first $k$ logarithmic shells falls off exponentially with $k$. A careful solution involves a renormalization group treatment of hierarchical resonances \cite{LevitovDipoles}, however the key scaling propeties can be read off from the analysis above. In particular, an excitation will almost surely find a site to hop to, ensuring that the state is {\it delocalized} on long lengthscales. %Nevertheless, transport on a scale $L$ is overwhelmingly likely to proceed via a single pairwise resonance at scale $L$, with matrix element $L^{-d}$, yielding (in three dimensions) the subdiffusive scaling $L \sim t^{1/3}$. 

Now let us adapt this argument to interacting systems %For specificity we consider a system in three dimensions, with $1/r^6$ interactions (which could be e.g. Van der Waals), although the arguments should carry through {\it mutatis mutandis} for a system in $d$ spatial dimensions 
with interactions that decay as $1/r^{2d}$. The prototypical Hamiltonian of interest takes the form

\begin{equation}
\hat H = \sum_{i} \epsilon_{i} S^z_{i} + \sum_{i \neq j} \frac{V}{|r_{i} - r_{j}|^{2d}} \big( S^{+}_i S^{-}_j + H.C. + J S^z_i S^z_j \big) \label{eq:H}
%H = \sum \epsilon_{\alpha} n_{\alpha} + \sum_{\langle \alpha \beta \rangle} \sum_{\langle \gamma \delta \rangle} \frac{V}{r^6_{\alpha \gamma}} b^{\dag}_{\alpha} b^{\dag}_{\gamma} b_{\delta} b_{\beta} \label{eq:Heff}
\end{equation}
%
%where the Greek subscripts refer to localized single particle eigenfunctions, we have thrown away purely diagonal interactions and assumed charge conservation, and $\alpha, \beta$ are nearest neighbors, as are $\gamma \delta$, and $r_{\alpha \gamma}$ represents the distance between the two pairs of nearest neighbors. 
Here the $\vec{S}_i$ are spin $1/2$ operators, which can be thought of as tracking whether a particular wavefunction (which is localized at the non-interacting level) is occupied or unoccupied, the $\epsilon_i$ are random numbers drawn from a distribution of width $W$, and $J$ is an $O(1)$ parameter, the precise value of which is unimportant for the analysis.  The $d=1$ version of this Hamiltonian is relevant for experiments with trapped ions \cite{trappedionrmp}, and the $d=3$ version is relevant for Van der Waals interactions e.g. in three dimensional Rydberg atom arrays \cite{Saffman}. I have deliberately adopted a notation that parallels \cite{YaoDipoles}. 

Let us assume that we are working close to a zero entropy state, with small but non-zero density of excitations $\rho$. This could correspond to working close to the ground state at low but non-zero temperature, or it could correspond to working close to a `fully polarized' state as in \cite{HoChoi}.  The key control parameter for our calculation will be 
\begin{equation}
\lambda = V \rho^2 /W \ll 1.
\end{equation}
Clearly, this control parameter can be tuned either by changing interaction strength, disorder strength or excitation density (which in turn could be altered by tuning temperature, if we were working close to the ground state, or by tuning magnetization, if we were working close to a fully polarized state). 

To formalize the calculation, I divide up the Hamiltonian as $\hat H_0 + \hat V$, where $\hat H_0$ is diagonal in the $Z$ basis, and $\hat V$ is off diagonal. Note that both $
\hat H_0$ and $\hat V$ have a long range $1/(r^{2d})$ tail. Now consider a pure state (but not necessarily eigenstate) initial condition $|\Psi\rangle$, which has density of excitations $\rho$, and work in the Schrodinger picture. It is convenient but not essential to consider $|\Psi\rangle$ to be a product state in the $Z$ basis, and to adopt a convention whereby `excitations' correspond to the system having local $S^z$ eigenvalue $+1/2$. The $\langle \Psi(t) | S^z_i| \Psi(t) \rangle$ are integrals of motion with respect to the Hamiltonian $\hat H_0$, but what happens in the presence of $\hat V$? I address this question within perturbation theory in small $V$. %We will not worry about non-perturbative effects such as avalanches \cite{avalanche}, since our analysis is concerned purely with {\it perturbative} stability. 

The perturbative analysis follows \cite{pwa}. The perturbation theory is structured in terms of matrix element numerators, and energy denominators. When the numerator is small compared to the denominator (off-resonance) this corresponds to {\it virtual} hopping and does not transport excitations. In contrast, {\it resonant} hops, where the matrix element equals or exceeds the energy denominator, {\it do} move excitations around. In principle, resonant hops can arise at any level in perturbation theory. However, it is by now well established that (for our model) in the limit $\rho \rightarrow 0$, when excitations are so dilute as to be effectively non-interacting, that resonances are rare and do not percolate, and that the problem is well localized at strong disorder \cite{pwa}. In contrast, at $\rho \neq 0$, a process first identified by Burin \cite{Burin} guarantees percolation of `resonant' rearrangements if the interactions fall off more slowly than $1/r^{2d}$. Our interactions, however, fall off {\it exactly} as $1/r^{2d}$, and are thus marginal with respect to the Burin criterion. 

Our analysis of marginally long range interactions proceeds by generalizing the analysis of \cite{Levitov89} for the non-interacting problem. Consider a logarithmic shell of radius $2^k R < r < 2^{k+1} R$. The volume of the $k^{th}$ region is $V_k \sim (2^k R)^d$. This contains $\rho V_k $ excitations. The characteristic level spacing for two particle states in this volume is $ \Delta_k = W / (\rho V_k)^2$. This sets the size of the typical energy denominators in perturbation theory. Meanwhile the matrix element on this lengthscale is $V/(2^k R)^{2d}$. The probability of a resonance in the $k^{th}$ logarithmic shell is thus equal to $P_k$, given by 
\begin{equation}
P_k = \frac{V/(2^k R)^{2d}}{ W / (\rho (2^k R)^d)^2} = \frac{V \rho^2}{W} = \lambda \ll 1 \label{eq:Pk}
\end{equation}
and is independent of $R$ and $k$. Now the probability of no resonance for any $k$ upto some macroscopic $N$ is $(1-\lambda)^N \rightarrow 0$, so resonances almost surely exist, with a broad distribution of lengthscales. When a resonance exists, we should re-diagonalize the problem to obtain new effective eigenstates, which will be bi-localized on two sites with separation $2^k R$, and with level splitting on the order of the matrix element, $V/(2^k R)^{2d}$. However, the existence of a sparse set of long range resonances is not in itself sufficient to produce delocalization - the resonances need to percolate \cite{pwa, Levitov89}, and I now argue that in our problem of interest, they do {\it not} percolate. 

To understand the failure of resonances to percolate, suppose you have $m$ resonances. The probability they all have the same lengthscale upto a factor of two is $\lambda^{m-1} \ll 1$. Thus, triples (and higher order resonances) are {\it rare} when $\lambda \ll 1$ - given two resonances, one of them will have a much smaller lengthscale than the other. The resonance with the smaller lengthscale will then develop a splitting that will be {\it large} compared to the matrix element of the longer-ranged resonance. It follows from the above that resonances cannot percolate. Suppose we are doing real space RG, and have coarse grained up to a scale $R$, whereupon a resonance first appears. This resonance will be an isolated resonance (by the above argument) and will have two body energy splitting equal to {\it at least} $V/R^{2d}$. Now, we are doing perturbation theory in $\hat V$, and at later stages we will be perturbing with an interaction $V/r^{2d}$ where $r \gg R$ by postulate.  This cannot be as large as the two body level splitting for this resonance. So this resonance cannot be involved again at a later stage of the RG. Thus, for critically long range interactions (in contrast to the case of critically long range hopping) {\it we do not expect resonances to percolate}. Resonances do indeed form at all scales, as in the case of critically long range hopping, but this time they are all {\it isolated} resonances, not percolating resonances, and thus they do not form a heat bath. As such, we expect the system to be {\it localized}, but critically so insofar as there exist isolated few body resonances on all lengthscales.

I clarify that the argument I have presented only establishes stability of localization at low orders in perturbation theory. Localization could still be destabilized at high orders, or by fully non-perturbative effects like the avalanche. I will discuss the associated timescales in due course. However, first I would like to discuss the phenomenology of the `critically localized' regime - I will show that this critically localized regime should persist all the way up to the avalanche timescale. 

I start by discussing some properties of the MBL regime which are common to all systems with interactions that decay as $1/r^{\alpha}$ with $\alpha \ge 2d$. These properties rely on the state being stable at low orders in perturbation theory, but do {\it not} rely on the stability being critical (i.e. $\alpha = 2d$). I start by noting that {\it most} sites are not part of any resonance. Thus, one could imagine `projecting' out the (non-percolating!) resonances above some cutoff lengthscale to obtain an effectively localized problem \cite{Geraedts}. Since most sites do not participate in any resonance, autocorrelation functions for a typical site should look indistinguishable from those of a conventional short range interacting localized phase \cite{NGH}. Meanwhile, the resonances that {\it do} exist are isolated, and thus should manifest as {\it sharp} spectral lines, e.g. in non-linear spectroscopy experiments \cite{armitage, NGdipoles} that couple to the relevant flip flop process. Since resonances exist on all lengthcales (and thus, on all energy scales), non-linear spectroscopy should see sharp resonant spectral lines all the way down to zero frequency, inside a localized phase. This is similar to the behavior that obtains in short range interacting systems. However, the existence of power laws also produces notable deviations from `short range interacting behavior.' Consider the dynamics of entanglement starting from unentangled (non-eigenstate) initial conditions. Entanglement entropy will grow with time as $t^{1/\alpha}$ instead of the conventional logarithmic in time growth for short range interacting systems, for reasons anticipated in \cite{Pino, Naini}, and will presumably saturate to volume law, much as it does in short range interacting localized systems. The `zone of disturbance' in nonlinear response \cite{KNS} will likewise scale as $\sim \tau^{1/{\alpha}}$ with the timescale $\tau$ on which the system is perturbed, instead of the logarithmic scaling obtained for short range interacting systems. The general theme is that various `logarithms' (in the short range interacting problem) get turned into power laws, which have the advantage of being easier to observe experimentally. These features are common to all power laws with $\alpha \ge 2d$. If $\alpha < 2d$ then localization is destabilized already at low orders in perturbation theory, and there is no MBL regime. 

I now turn to features that are particular to the `critical' power law ($\alpha = 2d$) that we have discussed herein. These properties will rely on the fact that low order resonances exist at {\it all} lengthscales, and moreover the density of resonances on lengthscale $R$ decays precisely as $R^{-d}$, so that resonances are equiprobable in {\it every logarithmic shell}. This is in contrast to conventional localized states, where the probability of resonances decays rapidly as one moves to larger logarithmic shells. The critical nature of the localization has distinctive signatures in the entanglement entropy of (approximate \footnote{Strictly speaking we are talking about approximate eigenstates - the true eigenstates are presumably thermal, but on timescales short compared to the avalanche timescale, the approximate eigenstates we are discussing herein will have no dynamics}) eigenstates. To zeroth order this should be area law (as in a typical localized state), but the long range resonances that straddle the entanglement cut will enhance the entanglement entropy. Since the resonances are uniformly and sparsely distributed in logarithmic distance, we will end up with an eigenstate entanglement entropy of $A \log L$, where $A$ is the area of the entanglement cut, and $L$ is the linear size of the smaller of the partitioned subregions in the direction normal to the cut. Such a logarithmic correction to the area law is familiar in e.g. Fermi liquid systems \cite{Klich, Swingle}, but here has completely distinct origins. %This provides an in-principle clean numerical diagnostic for the critical localization. 

A similar logarithm will arise in the statistics of the conserved charge ($S^z$), without the requirement to prepare the system in an (approximate) eigenstate. If one takes a critically localized system, not necessarily in an eigenstate, bipartitions the system, and measures the charge in one half of the system (e.g. in a quantum gas microscope), then the measured charge will have a quantum uncertainty associated with all resonances that straddle the boundary of the subregion, insofar as the charge measurement can `collapse' the resonance such that the charge shows up either inside or outside the subregion being probed. The number of such boundary straddling resonances will scale as $A \log L$, and so if the charge measurement is repeated multiple times to generate statistics, then the measured charge will be drawn from a distribution with standard deviation scaling as $A \log L$, and this logarithmic scaling provides an in principle experimental signature of critical localization which does not require the ability to prepare the system in eigenstates. 

Signatures also arise in frequency domain measurements of charge fluctuations. The conserved charge in a subregion will fluctuate due to resonances that span the boundaries of the subregion, and the {\it timescale} for the fluctuations will be determined by the characteristic energy scales of the resonances, which in turn are related to the lengthscales by $E \sim 1/R^{2d}$. Since the resonances are uniformly distributed in logarithmic lengthscale, they are also uniformly distributed in logarithmic energy scale. On converting to linear energy scale, we find that the probability distribution of resonances scales with frequency $f$ as $1/f$, and thus may be expected to produce $1/f$ noise, as another signature of the `critical' nature of the localization. 

Finally, energy transport will be subdiffusive. This last result follows from an argument similar to that employed in \cite{LevitovDipoles} - resonances exist at all lengthscales, and  energy can be transported on a lengthscale $L$ through a pairwise resonance on that lengthscale, but the matrix element falls off as $M_{fi} \sim 1/L^{2d}$, and the timescale may be extracted from $M_{fi} t \approx 1$ to yield the strongly subdiffusive scaling $L \approx t^{1/2d}.$ 
%{\bf Frequency dependent noise? Theory implications for critical points? Other manifestations?}

All the above phenomenology pertains to the MBL {\it regime} that is stable at low orders in perturbation theory. However, it is easy to see that this regime must be destabilized at high orders in perturbation theory. The key argument is adapted from \cite{tikhonovmirlin}, and is based on mapping the problem to \cite{agkl}. The mapping proceeds as follows: on a lengthscale $L$, there are $\rho^2 L^{2d}$ possible transitions that can be made, each with a matrix element $V/L^{-2d}$. One can then map this to a hopping problem on a tree \cite{agkl}, with parameters
\begin{equation}
K = \rho^2 L^{2d} \qquad Z = \frac{\rho^2}{\lambda} L^{2d},
\end{equation} where I have adopted the same notation as \cite{agkl}. An inspection of equation (12) from \cite{agkl} then leads to the conclusion that localization will necessarily be destabilized at order $n$ in perturbation theory if  
\begin{equation}
    f_n \approx \frac{\lambda}{\sqrt{n}} \left[ \lambda \log(\frac{\rho^2 L^{2d}}{\lambda n} ) \right]^{n-1} \ge 1,
\end{equation}
with the estimate being valid for $n \gg 1$. Now, at fixed large $n$ this leads to destabilizing $n^{th}$ order resonances at a lengthscale $L_n$ satisfying
\begin{equation}
    L_n^{2d} \approx \frac{\lambda n}{\rho^2} \exp\left( c_1 \frac{n^{1/(2(n-1))}}{\lambda^{n/(n-1)}} \right) \label{eq:Ln}
\end{equation}
where $c_1$ is an O(1) constant. This then yields a matrix element 
\begin{equation}
    M \approx \lambda^n L_n^{-2dn} %\approx \exp\left(-c_1 \frac{n}{\lambda^{n/(n-1)}}\right) 
    \label{matrixelement}
\end{equation}
where $L_n$ is given by Eq.\ref{eq:Ln}. Optimizing with respect to $n$ I obtain, at small $\lambda$, $n \approx \log(1/\lambda)\gg1$, consistent with our prior assumption that $n$ is large. Substituting back into Eq.\ref{matrixelement} I obtain (at small $\lambda$) the asymptotic timescale
\begin{equation}
    t_{agkl} \approx \exp \left(\tilde c_1 \frac{1}{\lambda} \log(1/\lambda)\right).
\end{equation}
where $\tilde c_1$ is another $O(1)$ numerical constant. This is the timescale on which effects at high order in perturbation theory may be expected to destabilize the localized regime. It should be contrasted with the avalanche timescale, which was estimated in \cite{gopalakrishnanhuse} (see their Eq. 18) as 
\begin{equation}
    t_{avalanche} \approx \exp(c_2 \log^2(\lambda)),
\end{equation}
where $c_2$ is an undetermined numerical constant. It is manifestly clear that at small $\lambda$ the avalanche timescale is parametrically {\it shorter} than the timescale from high-order breakdown, and thus it is the avalanche timescale that sets the limit of our MBL regime. 

To conclude: I have discussed the case of strongly disordered systems with critically long range two-body interactions falling off with distance as $1/r^{2d}$ in $d$ spatial dimensions. This problem is relevant to a range of experimental platforms, including Van der Waals interactions in three space dimensions and trapped ions in one space dimension. I have argued that, contrary to expectations, localization in such systems is perturbatively {\it stable}, but of critical character, with (non-percolating) resonances uniformly distributed in logarithmic lengthscale. This leads to a localized state with a distinctive phenomenology, including a `logarithmic correction' to the entanglement entropy of approximate eigenstates, and also a logarithmic scaling of the charge fluctuations (not necessarily in an exact eigenstate). This last provides an in principle experimentally accessible diagnostic which could be used to identify the critically localized regime in experiments. I have estimated the timescale on which this critically MBL regime is destabilized, and have argued that this is set by the avalanche timescale. %The physics we have discussed could be probed in multiple experimental platforms, including trapped ions in one dimension and three dimensional Rydberg atom arrays with Van der Waals interactions. 

{\bf Acknowledgements:} I thank Sarang Gopalakrishnan for various helpful discussions. This material is based upon work supported by the Air Force Office of Scientific Research under award number FA9550-20-1-0222. I also thank the Simons Foundation for support through a Simons Sabbatical Fellowship in Theoretical Physics.

\bibliography{LRMBL}
\end{document}